\documentclass[aps,prl,twocolumn,superscriptaddress]{revtex4}

\usepackage{mathrsfs,amssymb,graphics,subfigure,threeparttable}
\usepackage{grap hicx}
\usepackage{amssymb}
\usepackage{enumerate}
\usepackage{amsmath}
\usepackage{fancyhdr}
\usepackage{bm}
\usepackage[squaren]{SIunits}
\usepackage{xcolor}
\usepackage{changes}

\begin{document}
%\linenumbers
%\preprint{}

\title{Generation of terawatt, attosecond pulses from relativistic transition radiation}



%\author{Authors}
%\affiliation{SLAC National Accelerator Laboratory, Menlo Park, CA 94025}
%\affiliation{Department of Electrical Engineering, University of California, Los Angeles, California 90095, USA}
%\affiliation{Department of Physics and Astronomy, University of California Los Angeles, Los Angeles, CA 90095, USA}

\author{Xinlu Xu}
\email[]{xuxinlu@slac.stanford.edu}
\affiliation{SLAC National Accelerator Laboratory, Menlo Park, CA 94025}
\author{David B. Cesar}
\affiliation{SLAC National Accelerator Laboratory, Menlo Park, CA 94025}
\author{S\'ebastien Corde}
\affiliation{LOA, ENSTA Paris, CNRS, Ecole Polytechnique, Institut Polytechnique de Paris, 91762 Palaiseau, France}
\author{Vitaly Yakimenko}
\affiliation{SLAC National Accelerator Laboratory, Menlo Park, CA 94025}
\author{Mark J. Hogan}
\affiliation{SLAC National Accelerator Laboratory, Menlo Park, CA 94025}
\author{Chan Joshi}
\affiliation{Department of Electrical Engineering, University of California, Los Angeles, California 90095, USA}
\author{Agostino Marinelli}
\email[]{marinelli@slac.stanford.edu}
\affiliation{SLAC National Accelerator Laboratory, Menlo Park, CA 94025}
\author{Warren B. Mori}
\affiliation{Department of Electrical Engineering, University of California, Los Angeles, California 90095, USA}
\affiliation{Department of Physics and Astronomy, University of California Los Angeles, Los Angeles, CA 90095, USA}


\date{\today}

\begin{abstract} 
When a fs duration and hundreds of kA peak current electron beam traverses the vacuum and high-density plasma interface a new process, that we call relativistic transition radiation (R-TR) generates an intense $\sim100$ as pulse containing $\sim$ TW power of coherent VUV radiation accompanied by several smaller fs duration satellite pulses. This pulse inherits the radial polarization of the incident beam field and has a ring intensity distribution. This R-TR is emitted when the beam density is comparable to the plasma density and the spot size much larger than the plasma skin depth. Physically, it arises from the return current or backward relativistic motion of electrons starting just inside the plasma that Doppler up-shifts the emitted photons. The number of R-TR pulses is determined by the number of groups of plasma electrons that originate at different depths within the first plasma wake period and emit coherently before phase mixing. 
\end{abstract}

%\keywords{}

\maketitle

%\section{\label{sec: Introduction}Introduction}

%\begin{figure}[bp]
%\includegraphics[width=0.5\textwidth]{fig1.pdf}
%\caption{\label{fig: }  }
%\end{figure}

Transition radiation (TR) is a well-known phenomenon that happens when charged particles traverse an interface between two different media \cite{ginzburg1945radiation}. Electrons contained in the two media respond differently to the electric field carried by the charged particles and this leads to the emission of TR. An important case of TR happens when relativistic charged particles propagate across a vacuum-plasma/metal foil interface. In this scenario, electrons on the surface move in response to the transverse electric field of the beam, forming a surface current that screens the field and, in the process, emits radiation. For wavelengths longer than the bunch duration, the emitted photons are coherent, and the radiated pulse typically has a time structure similar to the bunch current profile. Therefore, coherent transition radiation is widely used to diagnose the longitudinal profile of (sub) picosecond electron beams \cite{happek1991observation, liu1998experimental, tremaine1998observation, lumpkin2001first, PhysRevLett.96.014801, glinec2007observation, lundh2013experimental, maxwell2013coherent}, as well as to produce intense THz radiation using kA current beams with subpicosecond duration \cite{leemans2003observation, daranciang2011single, wu2013intense}. If the bunch duration is shorter than the plasma oscillation period, then a wake is excited and a train of pulses will be generated \cite{hamster1993subpicosecond, sheng2004terahertz} at the plasma/foil-vacuum/neutral gas boundary after the initial transition radiation. %Each pulse will have a duration on the order of a plasma oscillation period with an amplitude that gradually decreases as phase mixing occurs.

\begin{figure}[bp]
\includegraphics[width=0.5\textwidth]{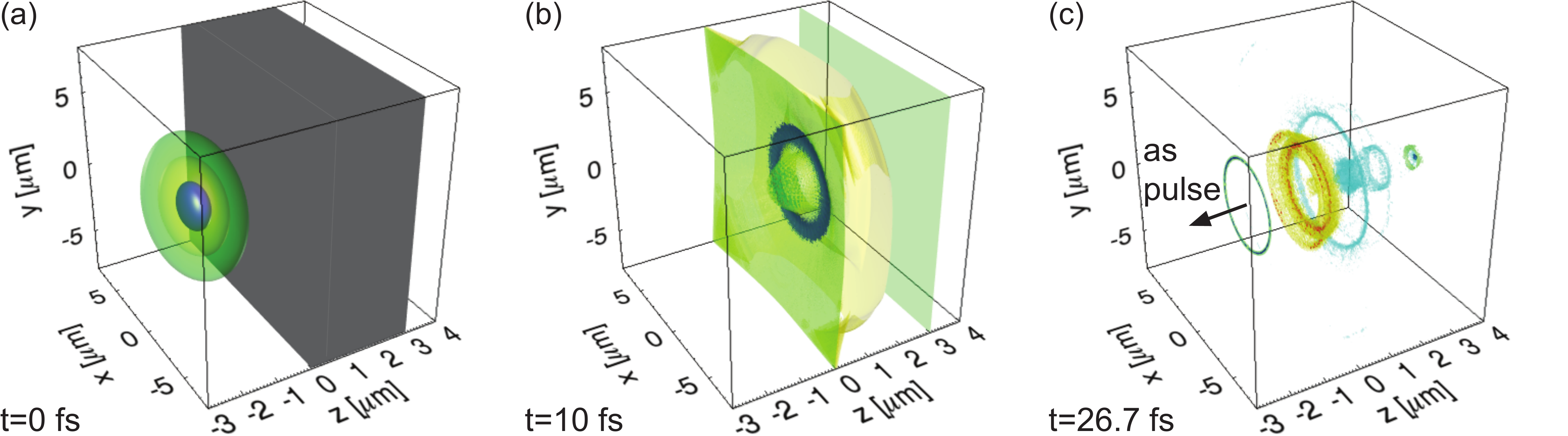}
\caption{\label{fig: } Relativistic transition radiation when an electron beam propagates through a plasma. (a) The charge density isosurfaces of the beam (blue, yellow and green) and the plasma (gray) at $t=0$fs. The beam center is $z=-1.5\micro\meter$. (b) The charge density isosurface of the plasma electrons at $t=10$fs. (c) The isosurfaces of the radiated transverse electric field $E_\perp$ (green and blue) and the axial current $j_z$ (red and yellow represent forward current density while cyan represents the backward current density) at $t=26.7$fs. }
\end{figure}

High brightness electron beams produced by state of the art photoinjectors can generate beams which have space-charge/Coulomb fields ranging from GV/m to TV/m \cite{blumenfeld2007energy, PhysRevLett.123.214801, duris2020tunable, yakimenko2019facet}, which can in turn accelerate free-electrons in the vicinity of the beam to relativistic energies during the transit time of the beam \cite{Supplement1}. However, even for such beam intensities, when the plasma density is greater than the beam density ($n_p\gg n_b$) the motion of electrons is non-relativistic due to the shielding effect of the free electrons and standard TR still occurs.

In this letter, we describe the new process of radiation emitted by intense beams close to the vacuum-plasma interface and call it relativistic transition radiation (R-TR). To understand the physics of R-TR it is useful to recall the normalized parameters that are used when describing plasma wakefield acceleration in the blowout regime \cite{PhysRevA.44.R6189, PhysRevLett.96.165002}. Consider a bi-Gaussian beam, $n_b=n_{b0}\mathrm{exp}(-\frac{r^2}{2\sigma_r^2}-\frac{z^2}{2\sigma_z^2})$, and the parameter $\Lambda\equiv \frac{n_{b0}}{n_p} (k_p\sigma_r)^2\equiv k_{pb}^2 \sigma_r^2$ where $k_{p,pb} \equiv \omega_{p,pb}/c$ and $\omega_{p,pb}$ is the plasma frequency using either the plasma or beam density respectively. In the relativistic blowout regime $\Lambda >1$ and the spot size of the beam is kept much less than a skin depth (equivalently when $n_{b0}\gg n_p$) and $k_p\sigma_z\sim 1$. Under these conditions the electrons move relativistically in response to the unshielded electric field of the beam.  A nonlinear wakefield is produced  and the currents from these wakes radiate from the boundary through a nonlinear mode conversion process which generates radiation at frequencies near the plasma frequency \cite{hamster1993subpicosecond, sheng2004terahertz}. However, a new process, R-TR, occurs if the spot size for the same beam (with $\Lambda\gg$ 10) is increased such that it is comparable to the blowout radius $2\sqrt{\Lambda} k_p^{-1}$ or equivalently when $n_{b0}\sim n_p$.  

%In this letter, we describe the process of TR emitted by intense beams whose physical picture is radically different than the traditional one and refer to this as relativistic transition radiation (R-TR). To understand the physics of R-TR it is useful to recall the normalized parameters that are used when describing plasma wakefield acceleration in the blowout regime \cite{PhysRevA.44.R6189, PhysRevLett.96.165002}. Consider a bi-Gaussian beam, $n_b=n_{b0}\mathrm{exp}(-\frac{r^2}{2\sigma_r^2}-\frac{z^2}{2\sigma_z^2})$, and the parameter $\Lambda\equiv \frac{n_{b0}}{n_p} (k_p\sigma_r)^2\equiv k_{pb}^2 \sigma_r^2$ where $k_{p,pb} \equiv \omega_{p,pb}/c$ and $\omega_{p,pb}$ is the plasma frequency using either the plasma or beam density respectively. In the relativistic blowout regime $\Lambda >1$ and the spot size of the beam is kept much less than a skin depth (equivalently when $n_{b0}\gg n_p$) and $k_p\sigma_z\sim 1$. Under these conditions the electrons move relativistically in response to the unshielded electric field of the beam. A nonlinear wakefield is produced  and the currents from these wakes radiate from the boundary through a nonlinear mode conversion process which generates radiation at frequencies near the plasma frequency. However,  we find in simulations that new process, R-TR, occurs if the spot size for the same beam (with $\Lambda\gg$ 10) is increased such that it is comparable to the blowout radius $2\sqrt{\Lambda} k_p^{-1}$ or equivalently when $n_{b0}\sim n_p$.  

Under these heretofore unexplored conditions, a large number of plasma electrons are pushed predominately forward rather than blown out which leads to a new regime of TR with multiple pulses. The first (we call zeroth) pulse has characteristics similar to traditional TR where the electric field is in the opposite direction to the beam's field, but its duration is longer than the bunch duration. However, there are subsequent pulses arising from the backward motion of electrons (return current now flows within the beam) arising deeper inside the plasma but from a region with thickness smaller than the wavelength of the first plasma wake. We find that making the target thicker so that it can support multiple wake periods has little effect on the emission of R-TR (supplemental movies M1-3). The most intense pulse comprises of photons that are Doppler-upshifted by the backward motion of the most energetic return current relativistic electrons. This results in radiation that has frequency components significantly higher than the natural plasma frequency. Particle-in-cell (PIC) simulations show one of the radiated pulses can be as short as 100 attoseconds with TW-level peak power. It can be generated by using femtosecond, ultra-high-current electron bunches envisioned for the next generation of plasma-wakefield experiments \cite{yakimenko2019facet}. This process is fundamentally different from the usual TR produced by the beam electrons as they traverse the plasma/vacuum interface.

Attosecond pulse generation has attracted much interest during the past two decades driven by its ability to resolve the electronic motion on the atomic scale \cite{PhysRevLett.68.1535, quere2006coherent, corkum2007attosecond, krausz2009attosecond, teubner2009high, thaury2010high, an2010enhanced, dromey2012coherent}. The R-TR process found here generates one high power attosecond pulse that is accompanied by several smaller pulses separated by $\sim 2\pi c/\omega_p$ due to the unique beam-plasma dynamics that also leads to some unique features such as radial polarization \cite{zhan2009cylindrical} and a narrow ring intensity distribution. The dominant attosecond pulse generated with this method could be synchronized with synchrotron or free-electron laser radiation emitted by the same electron bunch, therefore making possible X-ray pump-probe experiments with unprecedented accuracy. %Alternatively, one could separate the long-wavelength oscillations from the short attosecond burst with a frequency filter (or a thin foil), and combine them on a target for pump/probe or strong-field experiments.

% While most of this effort involves high-harmonic generation (HHG) near the tunnel ionization threshold of noble gases \cite{corkum2007attosecond}, high intensity attosecond pulses can be generated by interaction of high power (TW $\sim$ PW), ultra-short (10s of fs) laser pulses with solid density plasma ($10^{23}\sim10^{24}\centi\meter^{-3}$) \cite{teubner2009high, quere2006coherent, thaury2010high, an2010enhanced, dromey2012coherent}. The generation of R-TR is a three-dimensional (3D) effect while the aforementioned laser-driven schemes which generate usually a train of attosecond pulses via HHG can be essentially modeled using 1D simulations in a boosted frame \cite{PhysRevLett.68.1535, teubner2009high, thaury2010high}.

To demonstrate the physics behind R-TR and how it generates an isolated attosecond pulse, 3D fully relativistic PIC simulations are performed with the code OSIRIS \cite{fonseca2002high}. We use bi-Gaussian beams (defined above) and the simulations use normalized units. However, in what follows the actual beam parameters correspond to $6.5\times 10^9$  electrons and $\sigma_r=1.5\micro\meter$ and $\sigma_z=0.4\micro\meter$ to make connections to possible near-term experiments \cite{yakimenko2019facet}. This corresponds to $n_{b0}=3.7\times 10^{20} \centi\meter^{-3}$, a peak current of $I=250\mathrm{kA}$, and a skin depth for the beam density of $c/\omega_{pb} = 0.28\micro\meter$. As depicted in Fig. 1 (a), a 10GeV electron beam propagates into a plasma with immobile ions that exists between the $z=0\micro\meter$ and $z=3\micro\meter$. When the plasma density $n_p$ is much larger than the beam density $n_{b0}$, and $k_p\sigma_r\gg 1$, the space-charge field and current are  screened in a distance much less than the beam spot size so that well-known non-relativistic TR (non-R-TR) happens. This results in an electric field that opposes the beam's field and produces a radiative pulse with an electric field out of phase to the beam's vacuum field. In the opposite limit, where $n_p \ll n_{b0}$ and $k_p \sigma_r \ll 1$, the plasma electrons within a radius much larger than the beam spot size move relativistically outward and forward and form a narrow sheath \cite{PhysRevLett.96.165002}. These sheath electrons are then pulled backwards with relativistic energies, and this process is what sets up wakefields inside the plasma, provided the plasma is many skin depths long. As the electrons execute their full motion in the wake near the vaccum/plasma boundary they radiate. This radiation tends to be long wavelength and weak for wavelengths shorter than the plasma oscillation period. 
%We use a simulation box of size  $7 \micro\meter \times 100\micro\meter \times 100 \micro\meter$ with $2800\times1000\times 1000$ grids in the $z$, $x$ and $y$ directions, respectively. Each cell contains 1 simulation macro-particle.

\begin{figure}[bp]
\includegraphics[width=0.45\textwidth]{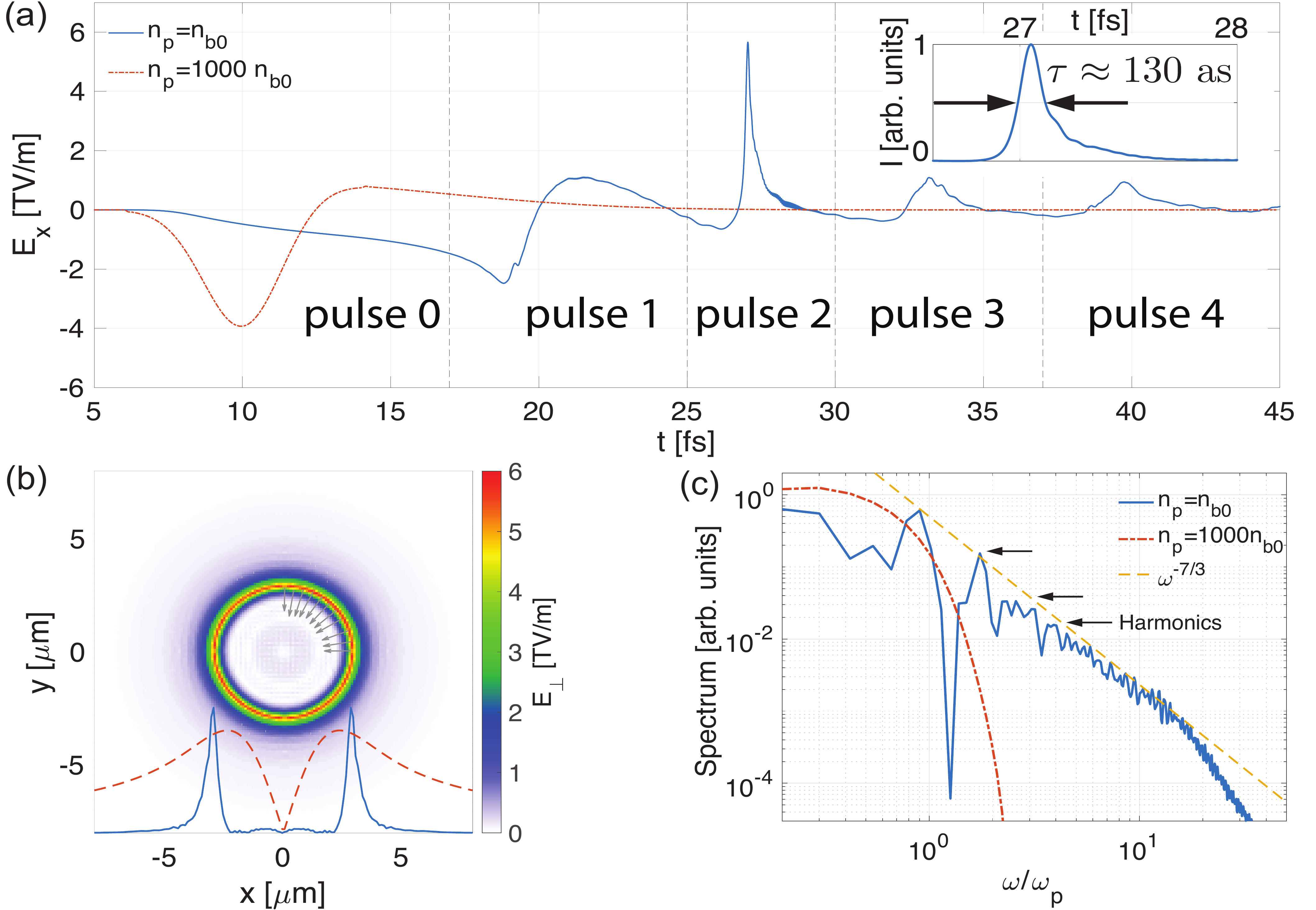}
\caption{\label{fig: } Attosecond radiation pulse. (a) The recorded electric field when $n_p=n_{b0}$ (blue) and $1000n_{b0}$ (red). Normalized intensity profile of the radiation (pulse 2) is shown in the inset. (b) The distribution of the transverse electrical field in the $x-y$ plane at $t=27.1\femto\second$ and $z=-1.5\micro\meter$. The gray arrows indicate the direction of the field. The solid blue line shows the field amplitude at $y=0\micro\meter$ and the dashed red line is for magnitude of the incident Coulomb field. (c) The spectrum of the radiation when $n_p=n_{b0}$ (blue), $1000n_{b0}$ (red) and the fitted scaling law (yellow).}
\end{figure}

However, when the plasma density and beam density are both high and roughly equal to one another (here a plasma with $n_p=n_{b0}=3.7\times 10^{20} \centi\meter^{-3}$ is used), plasma electrons that start inside and outside the beam can be accelerated to relativistic energies without a large transverse displacement. They are pushed outwards by the radial electric field of the beam and then gain forward velocity through the magnetic field, i.e., the $\bm{v}\times \bm{B}$ force. For the intense electron beams under investigation the plasma electrons are shot forward with more momentum than in the radial direction. For the bi-Gaussian charge distribution of the beam electrons,  plasma electrons initially at $r_i \sim 1.5\sigma_r$ experience the strongest space-charge field $E_r \approx cB_\theta \approx 0.45 \frac{n_{b0} e \sigma_r}{\epsilon_0}\mathrm{exp}\left(-\frac{z^2}{2\sigma_z^2}\right)$ and acquire the largest axial speeds, while electrons with $r_i \gg \sigma_{r}$ or $r_i \ll \sigma_r$ are  pushed forward less significantly. Thus, a region void of plasma electrons forms at the surface whose shape can be  approximated as a hollow cylinder, with a  annular cross section [Fig. 1(b)] in $r$ and a thickness $d$ (in $z$) is formed. The thickness can be estimated by balancing the forward force ($\frac{v_r}{c}B_\theta\sim E_r$) on plasma electrons from the beam with the charge separation force. If we further assume that $d \lesssim \sigma_r$ then the longitudinal field can be estimated from a one-dimensional argument. The thickness of the bare ions can thus be roughly estimated as $\frac{n_ped}{2\epsilon_0} \sim 0.45\frac{n_{b0} e \sigma_r}{\epsilon_0}$, i.e., $d\sim \frac{n_{b0}}{n_p} \sigma_r$, which is close to the $1~\micro\meter$ observed in simulations. As the plasma electrons move forward, a large space charge field is created at the surface through which the beam entered the plasma, and that field pulls backwards on electrons initially located deeper with the plasma. These electrons are then accelerated towards and across the boundary, emitting radiaiton we refer to as R-TR. Simulations show one intense attosecond pulse is emitted along the $-z$ direction which is compresed due to the relativistic Doppler effect [Fig. 1(c)]. There is some forward radiation as well but it is not as short nor intense as the backward one. Depending on the thickness of the target, this process can recur several times and produce several somewhat broader pulses with the energy in the later pulses progressively decreasing as phase mixing of electron oscillations occurs.

The detailed characteristics of the radiation from this R-TR are shown in Fig. 2. The time profiles of the transverse electric field at a fixed location $(z=-1.5\micro\meter, x=-2.95\micro\meter, y=0\micro\meter)$ are presented in Fig. 2(a) for R-TR ($n_p=n_{b0}$) and non-R-TR ($n_p=1000 n_{b0}$). The radiation $E_x$ emanating from the high-density plasma (red curve) is approximately proportional to the incident space-charge fields. In the R-TR case, there is a pulse with the electric field direction similar to the higher density case between 5 and 17fs for the R-TR (pulse 0, blue curve). The second (pulse 2, blue curve) and most prominent pulse has a half-cycle profile with very short, 130as,  duration (FWHM of the intensity) and a large peak electric field, 5.66TV/m, which is even larger than the incident space-charge field (4.5TV/m). The intensity profile of the pulse 2 is shown in the inset. The transverse cross section of the as pulse has an annular shape with 0.27$\micro\meter$ width (solid blue) which is much smaller than the 3.6$\micro\meter$ width of the space-charge field (dashed red). The radiation is radially polarized as indicated by the gray arrows in Fig. 2(b). The peak power is $1.47\tera\watt$ and the energy contained in pulse 2 is $0.29\milli\joule$. The transverse mode and the polarization of the radiation is inherited from the incident morphology of the space-charge field. Thus, a transversely asymmetric driver can be used to control the distribution and the polarization of the attosecond pulse \cite{Supplement2}.

\begin{figure}[bp]
\includegraphics[width=0.5\textwidth]{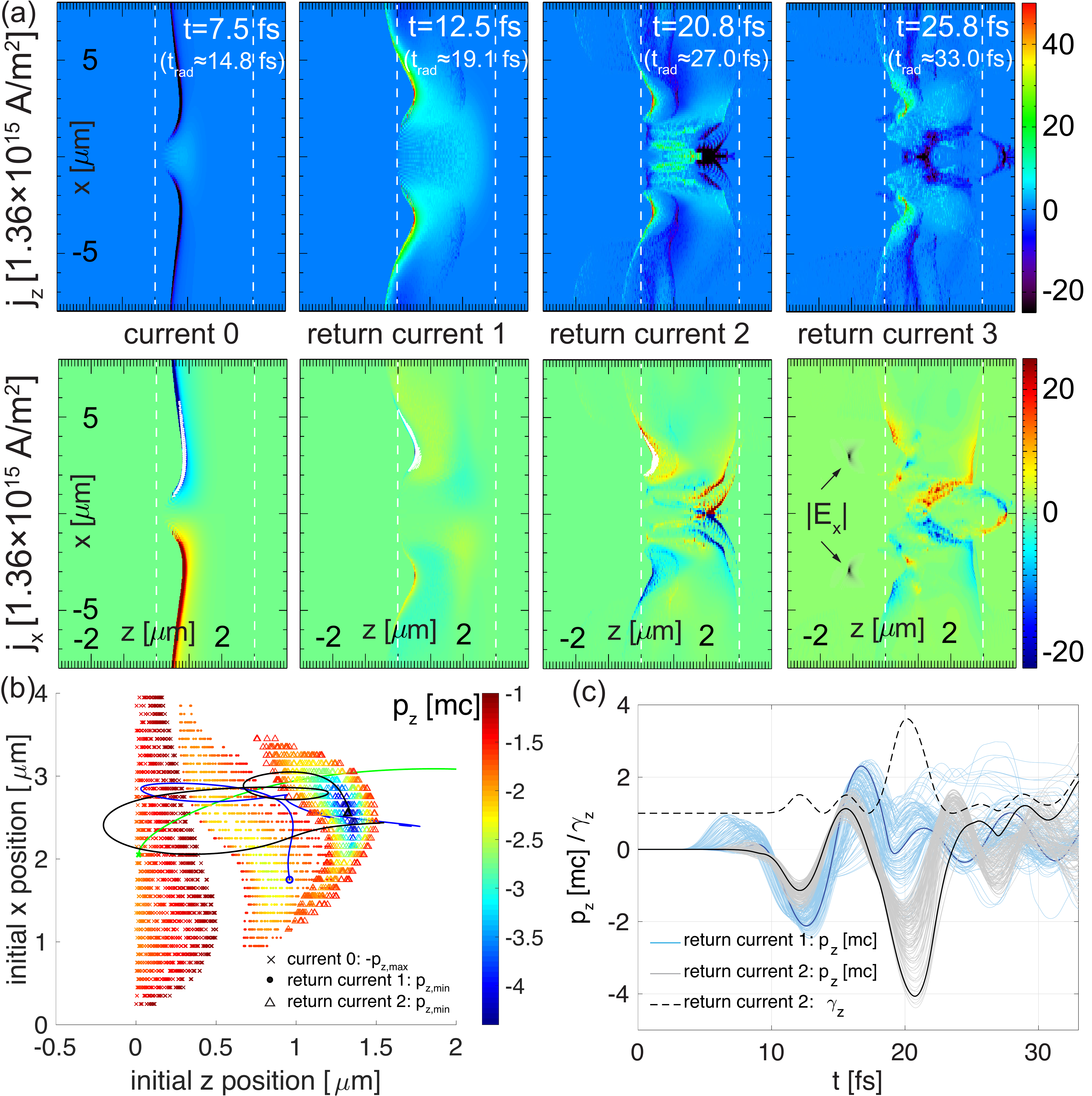}
\caption{\label{fig: } The motion of plasma electrons. (a) The $x-z$ plane distribution of the longitudinal ($j_z$) and transverse ($j_x$) current densities at four different times (t). The radiation generated at these times will arrive at the recorded position at $t_{\mathrm{rad}}$.  The dashed lines indicate the two boundaries of the target. The field distribution of the attosecond pulse (gray) is also shown in the last $j_x$ plot. (b) The initial positions of the electrons in three groups indicated by crosses, dots and triangles originating at different locations within the target determine the source of the current densities that radiates the zeroth, first and second EM pulse. The positions of each group of electrons when they coalesce into a high-density current sheath are shown as white dots in the corresponding $j_x$ plots of (a). Also shown are sample trajectories of three of these electrons (and supplemental movie M4-6). (c) The evolution of $p_z$ for the electrons in return currents. The solid lines represent two sample electrons and the black dashed line shows the $\gamma_z$ of one sample from return current 2. }
\end{figure}

We  compare the spectrum of the radiation from R-TR and non-R-TR in Fig. 2(c). The spectrum from non-R-TR is mainly from the incident beam that decays quickly (red dotted) after $\omega \sim c/\sigma_z\approx \omega_p$ while the one from R-TR extends to much higher frequencies because of the relativistic Doppler shift of the photons emitted by the backward moving electrons. This broad spectrum of the radiation extends up to $\omega\approx18\omega_p$ and provides the broad bandwidth needed to support the highest power, attosecond pulse. The central frequency of the photons in this ultrashort pulse is 7 eV which is in the VUV region. Harmonics of the nonlinear plasma oscillation are observed since one pulse is radiated from each plasma period. The ratio of the intensity between each peak shown in Fig. 2(a) is 0.19:0.038:1:0.037:0.028. Since the high frequency components are mainly contained in pulse 2, a high-pass filter can be used to enhance the intensity (contrast) ratio between pulse 2 and other pulses \cite{filter}. 

The details of the plasma response and surface currents generated by high current beams with $n_p=n_{b0}$ crossing a sharp interface are very complicated. This is illustrated by examining the current carried by the plasma electrons that start within $\sim2\pi k_p^{-1}$ (or 1.7 $\micro\meter$) the surface; these currents are the  source of the radiation and all the properties of the radiation can be traced back to the motion of these electrons. The longitudinal and transverse current distributions at four different times ($t_\mathrm{rad}$) when the R-TR pulses are emitted [Fig 2(a)] are shown in Fig. 3(a). Since electrons close to the surface emit the R-TR we identify a group of electrons that form the current (white dots in $j_x$ plots) and trace their trajectories backwards. The initial positions of each groups of these electrons are shown in Fig. 3(b). The first group [crosses in Fig. 3(b)] is called current 0 because these electrons are mostly pushed forward by beam. When the beam enters the plasma its radial electric field first pushes them outward and, once they acquire relativistic energies, forward by the magnetic field of the beam (supplemental movie M4) as represented by the trajectory of one representative electron (green) shown in this figure. Trajectory crossing of electrons from this group starting at different axial positions occurs and these electrons coalesce into similar axial positions leading to a high-density current sheet  as shown in first column of Fig. 3(a). The transverse components of these currents generate pulse 0 [5-17fs in Fig. 2(a)]. 
%which has an elongated profile compared with the incident space-charge field due to the forward relativistic motion of these electrons. 

After the driver passes, electrons that started deeper inside the plasma [dots in 3(b)], and that were overtaken by the first group of electrons closer to the surface, and hence accelerated by the fields of the beam less significantly, now move radially into the region of the strong fields within the hollow cylinder and are then accelerated backwards. Thus, a forward flowing current distribution (return current 1) is formed as shown in the second column of Fig. 3(a) and gives rise to the radiated pulse 1 [17-25fs in Fig. 2(a)]. These backward flowing electrons then cross into the vacuum. A sheath field is generated at the boundary that then pulls them back into the plasma (see the blue representative trajectory) where they are met by a group of electrons that started with axial positions slightly even deeper in the plasma [triangles in 3(b)]. The  forward moving  electrons pass through the second group of return current electrons, causing  the second group of electrons to be further accelerated backwards thus forming return current 2. This second group thus has higher energies than the first group as seen in their current distribution shown in the third column of Fig. 3(a). Intense attosecond pulse [25-30fs in Fig. 2(a)] is emitted by this high density return current sheath. This mixing of the trajectories of electrons that started at different axial positions leads to several bursts of the return current before the entire process phase mixes away. The return current formed by the third group is shown in the fourth column of Fig. 3(a). At this time, the radiated attosecond pulse has already moved away from the rear surface.

\begin{figure}[bp]
\includegraphics[width=0.5\textwidth]{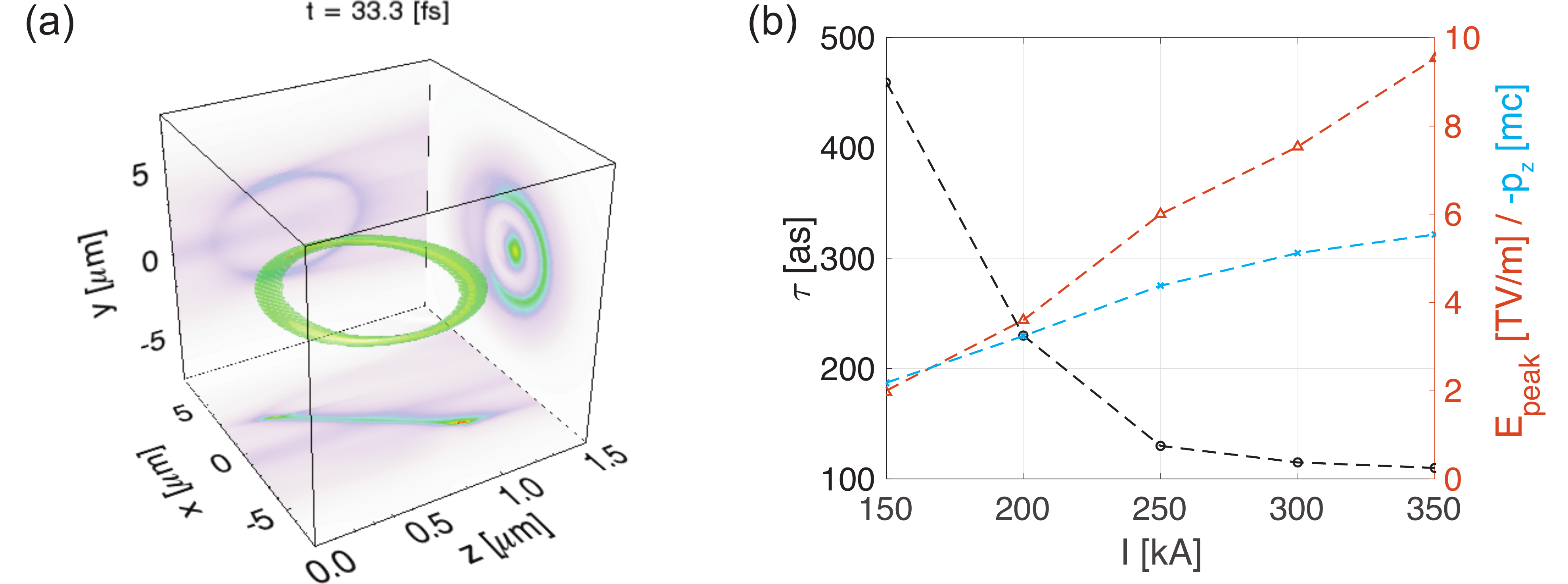}
\caption{\label{fig: } (a) The isosurface of the radiation intensity and its projections in each plane when the beam propagates along the $z$-direction and the target is oblique in the $z-x$ plane with angle $\theta=5.7\degree$. (b) The duration, peak electric field of the attosecond pulse and the maximum backward momentum of the electrons when scanning the current of the normally incident beam. }
\end{figure}

The axial momentum of the electrons in the return currents are shown in Fig. 3(c). The electrons in return current 1 (dots) achieve a maximum backward momentum $\sim 2mc$ around $t\sim12$fs. As a comparison, the electrons in return current 2 (triangles) achieve even larger backward momentum $\sim 4mc$ around $t\sim 20$fs. Compared with a relativistic plasma mirror, the electrons here only achieve the maximum backward momentum during a short time. And this `$\gamma$-spike' effect leads to a further reduction of the radiated pulse duration \cite{baeva2006theory}. A crude estimation of the radiation pulse duration can be given as $\tau \sim \frac{\tau_{\gamma_z}}{2\gamma_z^2}\approx  \frac{3~\mathrm{fs}}{2\times4^2}\approx 100\mathrm{as}$, where $\tau_{\gamma_z}$ is the duration of the `$\gamma$-spike'.

When the electron driver is  incident obliquely upon the target with a small angle, the radiation is produced approximately along the direction of specular reflection. The pulse produced by the return current 2 is shown in Fig. 4(a) for an angle of incidence of $\theta=5.7\degree$. In Fig. 4(b), we show how the maximum backward momentum of the electrons in return current 2 and the duration and the peak field of the attosecond pulse is affected by the peak current of the electron beam. When the peak density (and current)  of a beam with fixed charge (0.84nC) and 1.5$\micro\meter$ spot size is increased, the maximum backward momentum of electrons increases, and the duration of the attosecond pulse decreases and the field amplitude increases. The density of the plasma is also higher as it is always set equal to the peak density of the beam to ensure an efficient generation of attosecond pulse. The attosecond pulse generation is relatively insensitive to exact profiles of the plasma, its thickness and its initial temperature (supplemental information). Simulations with moving hydrogen or carbon ions give similar attosecond pulses (supplemental information).  

In conclusion, using 3D PIC simulations we have demonstrated that relativistic transition radiation is generated when a high current relativistic electron beam propagates through a plasma with $n_{b0}\sim n_p$. For realizable experimental parameters, the radially polarized radiation has a ring intensity distribution, is more intense in the backward direction and contains a $\sim$ terawatt,  $\sim$ 100 attosecond pulse flanked by several smaller fs pulses.

\begin{acknowledgments}
Work supported by the U.S. Department of Energy under contract number DE-AC02-76SF00515, the Department of Energy Basic Energy Sciences accelerator and detector research program, No. DE-SC0010064, and SciDAC FNAL subcontract 644405, and NSF Grants Nos. 1806046 and 1734315, and the European Research Council (ERC) under the European Union’s Horizon 2020 research and innovation programme (Miniature beam-driven Plasma Accelerators project, Grant Agreement No. 715807). The simulations were performed on the UCLA Hoffman 2 and Dawson 2 Clusters, and the resources of the National Energy Research Scientific Computing Center.
\end{acknowledgments}

\bibliography{refs_xinlu}

\begin{thebibliography}{34}
\bibitem[{fil()}]{filter}
\bibinfo{note}{For example, the intensity ratio is 0.033:0.006:1:0.01:0.006
  after a numerical filter with form $\mathrm{exp}[ - \left(
  \frac{\omega/\omega_p-50}{46} \right)^{10}]$ is used and the pulse 2 acquires
  a duration of 88 as and a peak electric field of 3.2 TV/m.}
\end{thebibliography}

\end{document}